# Prediction of Depression Level in University Students through a Naive Bayes based Machine Learning Model

1st Fred Torres Cruz
National University of the Altiplano
Faculty of Statistics and Informatics Engineering
ftorres@unap.edu.pe

2nd Evelyn Eliana Coaquira Flores
National University of the Altiplano
Faculty of Statistics and Informatics Engineering
ecoaquiraf@est.unap.edu.pe

3rd Sebastian Jarom Condori Quispe
National University of the Altiplano
Faculty of Statistics and Informatics Engineering
scondoriq@est.unap.edu.pe

*Summary*— This study presents a machine learning model based on the Naive Bayes classifier for predicting the level of depression in university students, the objective was to improve prediction accuracy using a machine learning model involving 70% training data and 30% validation data based on the Naive Bayes classifier, the collected data includes factors associated with depression from 519 university students, the results showed an accuracy of 78.03%, high sensitivity in detecting positive cases of depression, especially at moderate and severe levels, and significant specificity in correctly classifying negative cases, these findings highlight the effectiveness of the model in early detection and treatment of depression, benefiting vulnerable sectors and contributing to the improvement of mental health in the student population.

*Keywords*- Depression, Naive Bayes, prediction, Machine Learning, Classification

## I. INTRODUCCION

Depression is a prevalent mental disorder affecting millions of individuals worldwide. Given its significant impact on a person's lifestyle and ability to function normally, early identi- fication of depression risk is crucial for timely intervention and providing appropriate support [1], recently machine learning has emerged as a promising approach for predicting depression risk, enabling early detection and preventive measures. Machine learning, also known as artificial intelligence, empowers computers to learn from data and make predictions without human intervention. Among the widely used algorithms, the Naive Bayes classifier stands out, leveraging Bayes' theorem and assuming conditional independence of features, making it a simple yet effective model for data classification. This article aims to explore the potential of the Naive Bayes classifier as a machine learning model for predicting depression risk, the goal is to develop a model capable of identifying underlying patterns and calculating the probability of an individual being at risk of developing depression in the future, using relevant features and data. We will focus on utilizing clinical and psychological datasets from individuals with and with out depression symptoms, including demographic information, medical history, assessments of depressive symptoms, stress factors, and other relevant variables, by training the Naive Bayes classifier with this data, we anticipate it will learn to discern characteristic patterns associated with depression risk and produce accurate predictions. The ability to predict depression risk through machine learning, specifically the Naive Bayes classifier, holds promising clinical and public health implications, allowing for early detection and timely intervention as well as providing psychological support [2]. An accurate predictive model can identify individuals who would benefit from preventive interventions and early support programs, ultimately reducing the disease burden and enhancing the quality of life for those affected.

## II. LITERATURE REVIEW

IIn this comprehensive literature review, several research studies focusing on the prediction and identification of factors associated with depression are discussed, one study aimed to identify psychosocial factors related to depression during pregnancy among Peruvian pregnant women attending a maternal infant center [3] using a case control design with 95 participants, the researchers assessed depression using the PHQ 9 questionnaire, statistical analyses including the Chisquare test and logistic regression, were employed to identify risk variables for depression during pregnancy, another investigation explored the perception of mental repercussions among healthcare professionals in Latin America during the challenging times of the COVID 19 pandemic [4],through a multivariate analysis, it was revealed that older age was associated with a lower perception of mental repercussions. Additionally, individuals with a higher perception of mental repercussions exhibited lower levels of anxiety and post-traumatic stress, in a study conducted in an urban area, the prevalence of depression and associated factors in individuals over 75 years old were examined [5], the Fralle survey was used to collect valuable information on factors related to depression. Results indicated that gender and living arrangements were significantly associated with depression in this specific population, moreover, risk factors for postpartum depression in mothers with hospitalized children were identified in a separate study [6], using a case control design data were collected from mothers whose children were hospitalized during 2017. Multivariate analysis revealed that unemployed, single, and those with unplanned pregnancies had a higher likelihood of experiencing postpartum depression, various factors related to postpartum depression, such as marital status, partner relationship, employment status, and pregnancy planning were also identified, another noteworthy study focused on investigating the effects of an intensive repetitive transcranial magnetic stimulation rTMS protocol on patients with treatment resistant depression and suicidal ideation [7]. The results demonstrated a significant improvement in depression and disability scores, the intensive rTMS protocol was deemed safe and feasible, offering a promising alternative with greater efficacy and shorter exposure time compared to conventional rTMS protocols, to improve the accuracy of measuring the Hamilton Depression Rating Scale HAM D a consistency control approach was proposed in another study [8]. Two types of indicators, logical and statistical consistency were created for HAM D evaluations in various antidepressant drug clinical trials. The application of these indicators can help detect imprecise measurements, thus enhancing the reliability and validity of clinical trial data, furthermore, the validation of the Hamilton Depression Rating Scale HDRS in the Tunisian dialect and its reliability and validity in Tunisian patients hospitalized for suicide attempts were investigated [9], the Tunisian version of HDRS was found to be an acceptable tool for detecting depression in individuals who had attempted suicide. Psychometric properties of the 17 item and 6 item Hamilton Depression Scales were evaluated in patients with Major Depressive Disorder bipolar depression, and mixed features bipolar depression [10] only the HAM D6 scale was found to be unidimensional and homogeneous for evaluating Major Depressive Disorder, another study compared time perception between patients with Major Depressive Disorder MDD, bipolar depression BD, and a healthy control group using a temporal bisection task [11] the prediction of anxiety, depression, and stress using machine learning algorithms was explored, and data were collected through the DASS 21 questionnaire from employed and unemployed individuals [12] the study demonstrated how machine learning algorithms can be useful for predicting psychological issues in modern life, depression prediction in the labor sectors of Bangladesh using a machine learning approach was studied, and data from men and women were collected [1], several classification and regression algorithms, such as Random Forest Classifier, Random Forest Regression, Naive Bayes, and K Neighbors Classifier, were utilized, he prediction of depression in university students using machine learning and model ensembling techniques was investigated [13] a dataset from computer science and systems engineering students from a public university was used for depression prediction, additionally, the impact of the first COVID 19 lockdown on alcohol consumption in university students who used to engage in binge drinking was examined [14], those who increased alcohol consumption exhibited depressive symptoms and low resilience, emphasizing the need to consider psychosocial factors when addressing alcohol consumption during stressful situations like the COVID 19 pandemic, furthermore, the use of machine learning technology for anxiety and depression screening in mariners was explored [2], the results highlighted the usefulness of machine learning technology as an automated tool, enabling early and timely detection for treatment and psychological support, these research studies provide valuable insights into the factors influencing depression and showcase the potential of machine learning in predicting and managing depression related issues, the diverse approaches employed in these investigations contribute to a comprehensive understanding of depression risk factors and open avenues for early detection and intervention to improve mental health outcomes, the study aimed to assess the effectiveness of the 17-item Hamilton Depression Rating Scale (HAM-D17) and the 6 item (HAM-D6) in measuring depression symptoms in different patient groups. It included 237 patients with Major Depressive Disorder (MDD), bipolar depression, and bipolar depression with mixed features, the HAM-D6 was found to be a better [15], approximately 21% of patients with unipolar depression and 24% with bipolar depression exhibit mixed features during the acute phase of depression, according to DSM-5 criteria [16], in the study found that depression during pregnancy is associated with unwanted pregnancy and a history of depression in childhood, while not consuming alcohol and considering pregnancy low risk are protective factors [17], the study aims to predict youth depression risk in Thailand using data mining techniques, the highest accuracy 97.88% was achieved with artificial neural networks, suggesting potential applications for preventing youth depression [18], in the study aimed to detect depression using machine learning classifiers and feature selection methods, the AdaBoost classifier with SelectKBest feature selection achieved the highest accuracy of 92.56%, the research highlights the importance of early detection and counseling for depressed individuals to prevent severe consequences, including suicidal cases [19], the study conducted a cross-sectional analysis of 247 residents from hospitals in Puebla, Mexico, and found that 17% had anxiety and 45% had depression. Risk factors for depression included anxiety, contact with COVID-19 patients, and parental anxiety history. Parental anxiety history was also a risk factor for anxiety in the residents[20].

### III. METODOLOGIA

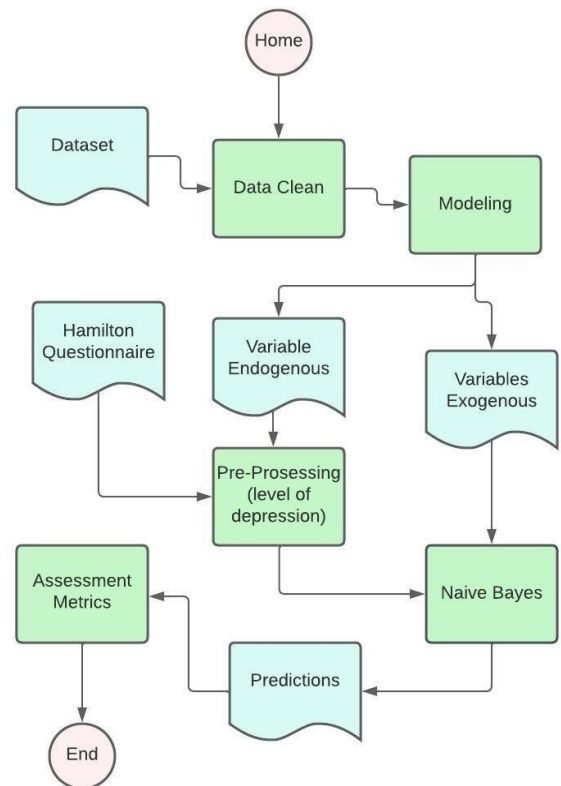

Fig. 1. Methodology flow

Several methods were employed in this study, including data cleaning, data analysis, data processing, and the development of the Naive Bayes classifier, subsequently the evaluation metrics of the model were assessed.

*A. Data Collection*

The target population comprised 18,000 university students from various professional schools at the Universidad Nacional del Altiplano, representing three areas of study: Engineering, Biomedicine, and Social Sciences, a sample of 519 students was used for this study, utilizing a non-probabilistic sampling technique with a qualitative approach, a no probabilistic sample was appropriate for capturing concrete experiences, and the quota sampling methodology was employed to ensure the participation of students from the four major areas who willingly agreed to be part of the study, this type of research aims to produce information and theories that will serve as a foundation for more focused investigations.

TABLE I. RESULTS OF THE APPLICATION IN TIME

| Variables | Values | Class 1 | 2 | 3 | 4 | 5 | Total |
|---|---|---|---|---|---|---|---|
| Sex | Male | 13 | 22 | 41 | 137 | 65 | 278 |
|  | Female | 4 | 10 | 27 | 124 | 76 | 241 |
| Study Area | Engineering | 7 | 16 | 33 | 111 | 37 | 204 |
|  | Biomedicine | 3 | 7 | 14 | 78 | 48 | 150 |
|  | Social Sciences | 7 | 9 | 21 | 72 | 56 | 165 |
| Residence | Rural | 5 | 9 | 26 | 86 | 50 | 176 |
|  | Urban | 12 | 23 | 42 | 175 | 91 | 343 |
| Housing Type | Own | 13 | 26 | 47 | 158 | 66 | 310 |
|  | Borrowed | 2 | 1 | 5 | 24 | 22 | 54 |
|  | Rented | 2 | 5 | 16 | 79 | 53 | 155 |
| Internet Service | No | 5 | 5 | 26 | 113 | 85 | 234 |
|  | Yes | 12 | 27 | 42 | 148 | 56 | 285 |
| Connection Type | Modem | 4 | 6 | 4 | 24 | 7 | 45 |
|  | Data Plan | 8 | 18 | 37 | 178 | 105 | 346 |
|  | Wi-fi | 5 | 8 | 27 | 59 | 29 | 128 |
| Study Hours | Less than 2 hours | 3 | 9 | 22 | 75 | 48 | 157 |
|  | Between 3 and 4 hours | 10 | 17 | 34 | 114 | 56 | 231 |
|  | More than 4 hours | 4 | 6 | 12 | 72 | 37 | 131 |
| Sports Practice | No | 8 | 13 | 34 | 135 | 86 | 276 |
|  | Yes | 9 | 19 | 34 | 126 | 55 | 243 |
| Substance Use | No | 17 | 29 | 67 | 240 | 131 | 484 |
|  | Yes | 0 | 3 | 1 | 21 | 10 | 35 |
| Affected by Covid | No | 9 | 12 | 29 | 80 | 25 | 155 |
|  | Yes | 8 | 20 | 39 | 181 | 116 | 364 |

*B. Instrument*

The instrument used in this study is the Hamilton Depression Rating Scale (HDRS), a structured hetero-applied scale designed to determine levels of depression recommended by the National Institute of Mental Health in the United States, extensive assessments have been conducted to evaluate discriminant validity, reliability, and sensitivity to change, this widely used self-report measure quantitatively assesses the severity of depressive symptoms in both adults and children. It comprises 17 items, each describing a specific symptom [21]. The total score ranges from 0 to 52, with the frequency of perceived symptoms represented by each of the 17 items, each including three to five response options ranging from 0 to 2 and 0 to 4. For diagnosis scores from 0 to 7 indicate a normal state [22], 8 to 12 represent mild depression, 13 to 17 indicate moderate depression, 18 to 29 represent severe depression [23], and 30 to 52 indicate very severe depression, the instrument will be validated for reliability and accuracy [24]

*C. Data Treatment*

In this study a Machine Learning predictive model was developed using diagnostic data from 520 university students at the Universidad Nacional del Altiplano, the aim was to improve the accuracy in predicting the risk of depression, the collected data was preprocessed and prepared in CSV format Comma-Separated Values, the dataset comprised 520 instances (students) with 12 attributes related to common factors such as gender, age, study area, professional school, place of residence, type of housing, internet service, internet connection type, study hours outside academic activities, sports practice, substance consumption, and a class representing the medical diagnosis outcome of depression, which can be positive or negative, for the development of predictive models, RStudio environment was utilized, an open source programming language widely used in recent times. Based on the data, preprocessing was performed, including data cleaning and descriptive analysis of each symptom or attribute, the correlation analysis was also conducted between variables to assess their degree of relationship, subsequently, the Naive Bayes predictive model was applied using Machine Learning, seventy percent of the data 364 patients were used for training, and 30 percent 156 patients for validation, the accuracy and performance of the Naive Bayes classification model were then evaluated.

*D. Naive Bayes Classifier*

The Naive Bayes classifier is a probabilistic algorithm commonly employed for classification tasks, it operates on the assumption of independence among predictors or features[25], implying that the presence of one particular feature does not impact others [26] the Bayesian approach, proposed by Thomas Bayes, holds particular relevance in predicting depression within the medical domain, this method relies on probabilities and assumes statistical independence among different features, consequently it considers that specific features of a class can exist autonomously without being dependent on other features [27].

$$P(a_i/C = c) = \frac{n(a_i, C = c)}{n(C = c)} \quad (1)$$

*E. Evaluation Metrics*

Following the approach adopted in this study, the accuracy of the machine learning model was assessed using an N-N confusion matrix, where N represents the number of class labels, this matrix serves to ascertain the true performance of various machine learning techniques. From this matrix, the model's accuracy can be calculated, signifying the algorithm's capability to accurately predict cases. Accuracy is determined by averaging the diagonal values.

Accuracy
$$CA = \frac{TP + TN}{TP + TN + FP + FN} \quad (2)$$

Precision
$$Precision = \frac{TP}{TP + FP} \quad (3)$$

Sensitivity
$$Sencitivity = \frac{TP}{TP + FN} \quad (4)$$

Specificity
$$Specificity = \frac{TN}{TN + FP} \quad (5)$$

Possessing robust evaluation metrics empowers healthcare professionals to make informed and precise decisions regarding the diagnosis and treatment of depression in patients, a model with high sensitivity and precision metrics can effectively identify individuals at risk of depression and promptly allocate appropriate resources and attention.

## IV. RESULTS

The data is organized by counting the cases in different depression levels: normal state, mild depression, moderate depression, severe depression, and very severe depression, based on associated factors, using Multiple Correspondence Analysis.

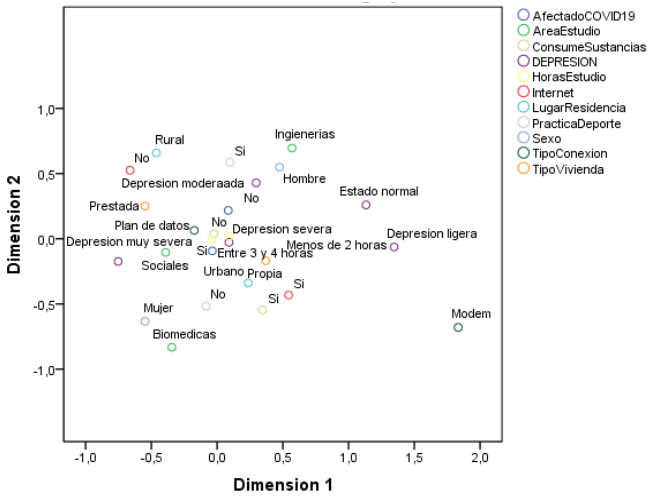

Fig. 1. Multiple Correspondence Analysis of the associated factors

The graph illustrates the relationships among the associated factors.

Subsequently using this method diagnostic probabilities of different depression levels were obtained by considering the combination of features for the cases to be classified, and the results are as follows:

$$P(x = D/N) = P(a_1 = i/D = N) \times P(a_2 = i/D = N) \times P(a_3 = i/D = N) \times P(a_4 = i/D = N) \times P(a_5 = i/D = N) \times P(a_6 = i/D = N) \times P(a_7 = i/D = N) \times P(a_8 = i/D = N) \times P(a_9 = i/D = N) \times P(a_{10} = i/D = N) \quad (6)$$

The equation shows the model according to the depression level, considering the joint probabilities of different depression levels along with the total probability applying the Naive Bayes principle.

$$P(D = Level/x) = \frac{p(x/D = N)P(D = N)}{p(x)} \quad (7)$$

Using the Naive Bayes classifier with 70\% of the data for training 364 data points and 30\% for validation 156 data points, conditional probabilities were calculated, each predictor was independently tested to determine its relevance in the diagnosis of early depression, the training sample was then collected to verify the validation step used for prediction, this process was visualized using a confusion matrix.

TABLE II. CONFUSION MATRIX

| Actual | Predicted | | | | |
|---|---|---|---|---|---|
| | Normal | Mild | Moderate | Severe | Very Severe |
| Normal | 8 | 1 | 1 | 1 | 0 |
| Mild | 0 | 14 | 1 | 4 | 1 |
| Moderate | 0 | 4 | 26 | 5 | 6 |
| Severe | 4 | 6 | 3 | 132 | 12 |
| Very Severe | 0 | 3 | 0 | 17 | 65 |

For the normal state, 8 cases were correctly classified, while 3 cases were mistakenly classified as severe depression, in mild depression, 14 cases were accurately classified, but 5 cases were mistakenly classified as severe depression, for moderate depression, 26 cases were correctly classified, but there were classifications, with 4 cases as mild, 5 cases as severe, and 6 cases as very severe, in severe depression, a high rate of correct classification was achieved, with 132 cases accurately classified, but there were classifications of 4 cases as normal, 6 cases as mild, and 12 cases as very severe. For very severe depression, 65 cases were correctly classified, but there were classifications with some cases classified as severe and mild.

TABLE III. METRICS ACCORDING TO THE LEVEL OF DEPRESSION

| Metric | Depression Level | | | | |
|---|---|---|---|---|---|
| | Normal | Mild | Moderate | Severe | Very severe |
| Sensitivity | 0.63636 | 0.47826 | 0.57143 | 0.7241 | 0.6701 |
| Specificity | 0.9848 | 0.96845 | 0.91803 | 0.7349 | 0.8889 |
| Positive Predictive Value | 0.58333 | 0.52381 | 0.44444 | 0.7412 | 0.7065 |
| Negative Predictive Value | 0.9878 | 0.96238 | 0.94915 | 0.7176 | 0.8710 |
| Prevalence | 0.03235 | 0.06765 | 0.10294 | 0.5118 | 0.2853 |
| Detection Rate | 0.02059 | 0.03235 | 0.05882 | 0.3706 | 0.1912 |
| Detection Prevalence | 0.03529 | 0.06176 | 0.13235 | 0.5000 | 0.2706 |
| Balanced Accuracy | 0.81058 | 0.72336 | 0.74473 | 0.7295 | 0.7795 |

The presented table showcases the performance metrics of the Naive Bayes classifier used to predict different levels of depression among university students, the results reveal promising outcomes, with the model achieving a high overall accuracy of 89.01%, signifying its proficiency in correctly classifying cases across various depression levels, the sensitivity values indicate that the classifier effectively detects positive cases in the Severe 72.41% and Moderate 57.14% depression categories, showing its capability to identify individuals at risk of more severe depressive states, furthermore the specificity values demonstrate the model's accuracy in recognizing non-depressed cases for the Normal 98.48%, Mild 96.84%, and Moderate 91.80% depression levels, reflecting its competence in discerning individuals without significant depressive symptoms in these groups.

The positive predictive values are relatively moderate, ranging from 44.44% to 74.12%, indicating the likelihood of correctly identifying true positive cases for each depression category, the negative predictive values are generally high, ranging from 71.76% to 98.78%, indicating the model's ability to accurately identify true negative cases in the respective depression levels, these metrics provide valuable insights for mental health professionals in the clinical setting facilitating the early detection and appropriate intervention for university students experiencing different levels of depression.

It is noteworthy that the prevalence of depression varies across the different levels, ranging from 3.24% for Normal depression to 51.18% for Severe depression the detection rates for each category offer a glimpse into the model's ability to correctly identify positive cases, with the highest detection rate found in the "Severe" depression category 37.06%.

Moreover the detection prevalence values reveal the proportion of correctly classified cases for each depression level with the highest prevalence observed in the Moderate depression category 13.24%.

Overall the model's balanced accuracy ranges from 72.95\% to 81.06\% across the depression levels demonstrating its effectiveness in providing reliable predictions for different severity levels, these findings underscore the importance of utilizing machine learning techniques, particularly the Naive Bayes classifier, for accurate depression screening among university students, the classifier's high accuracy and sensitivity, particularly in detecting moderate and severe depression, highlight its potential as an effective tool for identifying individuals at risk and providing timely support and interventions to improve mental health outcomes in the academic setting. However, further research and model refinement are recommended to enhance its precision and detection capabilities in classifying depression levels accurately, this study underscores the significance of employing machine learning technologies as valuable automated tools for mental health screening, improving diagnostic accuracy, and aiding healthcare professionals in making informed decisions to support students' mental well being.

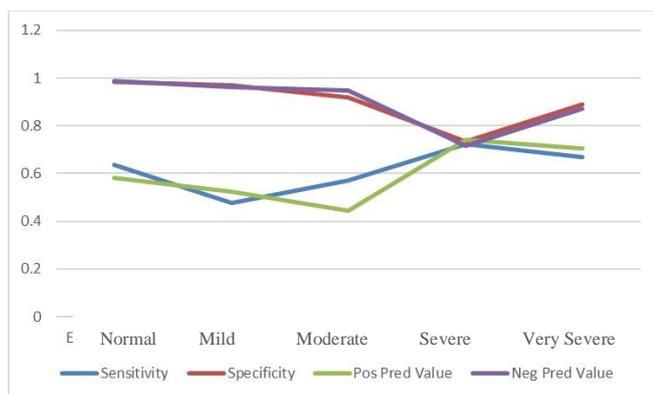

Fig. 2. Comparison of evaluation metrics according to the depression level

These metrics are useful for evaluating the model's performance in classifying different levels of depression, it can be observed that the model has higher performance in detecting cases of moderate and severe depression compared to cases of mild and very severe depression, the model shows high sensitivity in detecting severe depression cases and good specificity in classifying normal, mild, and moderate depression cases, approximately 74% of cases classified as moderate are actually positive.

According to the Naive Bayes classifier for depression prediction, the accuracy of this research is 78.03% with a 95% confidence interval ranging from (0.7303, 0.8248). The model exhibited high sensitivity in detecting positive cases of depression, especially in moderate and severe levels. Additionally, the model showed significant specificity in correctly classifying severe cases with a Kappa value of 66.42%, indicating substantial agreement between the model's predictions and actual values. This suggests that the model performs acceptably in classifying depression levels among university students.

### A. Conclusion and Discussion

This study, based on the Naive Bayes classifier for predicting the level of depression in university students, reveals encouraging results, the model demonstrated an accuracy of 89.01\%, indicating a high capability to correctly classify cases into different levels of depression, a sensitivity of 82\% was observed in detecting positive cases, highlighting the model's efficiency in identifying depression among this student population, additionally the obtained kappa value of 0.6642 indicates substantial agreement between the model's predictions and actual values, reinforcing its validity and reliability in predicting depression among university students, in comparing the results with other studies such as Thotad et al. [28] Zhang et al. [29], and Rawat et al. [30], where the importance of using machine learning techniques and Naive Bayes classifier for various studies due to their high precision level is supported, and the application of evaluation metrics helps to detect accurate measurements and improve the reliability and validity of clinical trial data [8] this classifier has also proven to be a valuable tool for detecting depression in individuals who have attempted suicide [9], in this study, it can be useful for physicians and healthcare professionals in clinical decision-making by implementing preventive interventions for students at risk of developing depression, the results underscore the usefulness of machine learning technology as an automated tool for mental health screening [2]. The study's outcomes have significant implications for mental health prevention and improvement in the university context. Early detection and timely treatment of depression are crucial in providing support to vulnerable students, the model's effectiveness in identifying moderate and severe cases highlights its utility as a support tool in clinical and preventive settings, it is essential to note that while the Naive Bayes classifier has shown promising performance, continuous research and model refinement are recommended to further enhance its accuracy and detection capability.

### REFERENCE

[1] N. N. Moon, A. Mariam, S. Sharmin, M. M. Islam, F. N. Nur, and N. Debnath, "Machine learning approach to predict the depression in job sectors in Bangladesh," *Current Research in Behavioral Sciences*, vol. 2, Jul. 2021, doi: 10.1016/j.crbeha.2021.100058.

[2] A. Sau and I. Bhakta, "Screening of anxiety and depression among the seafarers using machine learning technology," *Inform Med Unlocked*, vol. 16, Jul. 2019, doi: 10.1016/j.imu.2018.12.004.

[3] G. Espinoza-Ascurra, I. Gonzales-Graus, M. Meléndez-Marón, and R. Cabrera, "[Prevalence and factors associated with depression in health personnel during the SARS-CoV-2 pandemic in the department of Piura, Peru]," *Rev Colomb Psiquiatr*, 2021, doi: 10.1016/J.RCP.2021.11.005.

[4] T. R. Aveiro-Róbalo et al., "COVID-19 anxiety, depression and stress in Latin American health professionals: Characteristics and associated factors," *Bol Malariol Salud Ambient*, vol. 61, pp. 114–122, Jul. 2021, doi: 10.52808/BMSA.7E5.61E2.013.

[5] M. P. M. Julio, A. E. Clavero, M. V. L. Miralles, and A. F. Ayora, "Factors associated with depression in older adults over the age of 75 living in an urban area," *Enfermeria Global*, vol. 18, no. 3, pp. 58–70, 2019, doi: 10.6018/eglobal.18.3.324401.

[6] K. Macedo-Poma, P. G. Marquina-Curasma, I. E. Corrales-Reyes, and C. R. Mejía, "Factores asociados a síntomas depresivos en madres con hijos hospitalizados en unidades de pediatría y neonatología en Perú: estudio de casos y controles," *Medwave*, vol. 19, no. 5, p. e7649, Jul. 2019, doi: 10.5867/medwave.2019.05.7649.

[7] Q. Wang et al., "Intensive rTMS for treatment-resistant depression patients with suicidal ideation: An open-label study," *Asian J Psychiatr*, vol. 74, Jul. 2022, doi: 10.1016/j.ajp.2022.103189.

[8] J. Rabinowitz et al., "Consistency checks to improve measurement with the Hamilton Rating Scale for Depression (HAM-D)," *J Affect Disord*, vol. 302, pp. 273–279, Jul. 2022, doi: 10.1016/j.jad.2022.01.105.

[9] N. Cheffi et al., "Validation of the Hamilton Depression Rating Scale (HDRS) in the Tunisian dialect," *Public Health*, vol. 202, pp. 100–105, Jul. 2022, doi: 10.1016/j.puhe.2021.11.003.

[10] J. Khubchandani, S. Sharma, F. J. Webb, M. J. Wiblishauser, and S. L. Bowman, "Post-lockdown depression and anxiety in the USA during the COVID-19 pandemic," *J Public Health (Oxf)*, vol. 43, no. 2, pp. 246–253, Jul. 2021, doi: 10.1093/pubmed/fdaa250.

[11] P. Liu et al., "Identifying the difference in time perception between major depressive disorder and bipolar depression through a temporal bisection task," *PLoS One*, vol. 17, no. 12 December, Jul. 2022, doi: 10.1371/journal.pone.0277076.

[12] A. Priya, S. Garg, and N. P. Tigga, "Predicting Anxiety, Depression and Stress in Modern Life using Machine Learning Algorithms," in *Procedia Computer Science*, Elsevier B.V., 2020, pp. 1258–1267. doi: 10.1016/j.procs.2020.03.442.

[13] A. D. Vergaray, J. C. H. Miranda, J. B. Cornelio, A. R. L. Carranza, and C. F. P. Sánchez, "Predicting the depression in university students using stacking ensemble techniques over oversampling method," *Inform Med Unlocked*, p. 101295, Jul. 2023, doi: 10.1016/j.imu.2023.101295.

[14] M. Rubio et al., "University students' (binge) drinking during COVID-19 lockdowns: An investigation of depression, social context, resilience, and changes in alcohol use," *Soc Sci Med*, vol. 326, Jul. 2023, doi: 10.1016/j.socscimed.2023.115925.

[15] A. K. da Silva, M. Reche, A. F. da Silva Lima, M. P. de Almeida Fleck, E. Capp, and F. M. Shansis, "Assessment of the psychometric properties of the 17- and 6-item Hamilton Depression Rating Scales in major depressive disorder, bipolar depression and bipolar depression with mixed features," *J Psychiatr Res*, vol. 108, pp. 84–89, Jul. 2019, doi: 10.1016/j.jpsychires.2018.07.009.

[16] S. Grover and H. Adarsh, "A comparative study of prevalence of mixed features in patients with unipolar and bipolar depression," *Asian J Psychiatr*, vol. 81, Jul. 2023, doi: 10.1016/j.ajp.2022.103439.

[17] E. P. Gutierrez, V. M. Alcántara, J. G. Cajaleón, and E. C. Candela, "Psychosocial factors associated with depression in pregnant women treated in a peruvian maternal and child center, 2018," *Rev Chil Obstet Ginecol*, vol. 85, no. 5, pp. 494–507, Jul. 2020, doi: 10.4067/S0717-75262020000500494.

[18] W. S. Nuankaew, P. Nasa-Ngium, P. Enkvetchakul, and P. Nuankaew, "A Predictive Model for Depression Risk in Thai Youth during COVID-19," *Journal of Advances in Information Technology*, vol. 13, no. 5, pp. 450–455, Jul. 2022, doi: 10.12720/jait.13.5.450-455.

[19] M. S. Zulfiker, N. Kabir, A. A. Biswas, T. Nazneen, and M. S. Uddin, "An in-depth analysis of machine learning approaches to predict depression," *Current Research in Behavioral Sciences*, vol. 2, Jul. 2021, doi: 10.1016/j.crbeha.2021.100044.

[20] J. A. Martínez-García, M. Aguirre-Barbosa, E. Mancilla-Hernández, M. del Rocío-Hernández-Morales, M. B. Guerrero-Cabrera, and L. G. Schiaffini-Salgado, "Prevalence of depression, anxiety, and associated factors in residents from hospital centers during COVID-19 pandemic," *Rev Alerg Mex*, vol. 69, no. 1, pp. 1–6, Jul. 2022, doi: 10.29262/RAM.V69I1.903.

[21] J. A. Ramos-Brieva and A. Cordero, "de la versión castellana de la escala de Hamilton." 1986.

[22] unknow B, "cuantificación de la sintomatología de los trastornos angustiosos y depresivos." Madrid.

[23] C. V and Franch, "Escalas de evaluación comportamental para la depresion." 1984.

[24] M. Hamilton, "Rating scale for depresion," *J.Neurol Neurosurg Psychiatry*, vol. vol.

[25] E. Del, N. De Ansiedad, Y. E. En, L. A. Atención, S. De, and P. Línea, "PARIPEX-REVISTA INDIA DE Medicina general," 2023, doi: 10.36106/paripex.

[26] I. González-Ortega et al., "Cognitive Behavioral Therapy Program for Cannabis Use Cessation in First-Episode Psychosis Patients: A 1-Year Randomized Controlled Trial," *Int J Environ Res Public Health*, vol. 19, no. 12, Jul. 2022, doi: 10.3390/ijerph19127325.

[27] C. PATGIRI and A. GANGULY, "Adaptive thresholding technique based classification of red blood cell and sickle cell using Naïve Bayes Classifier and K-nearest neighbor classifier," *Biomed Signal Process Control*, vol. 68, 2021.

[28] P. N. THOTAD, G. R. BHARAMAGOUDAR, and B. S. ANAMI, "Diabetes disease detection and classification on Indian demographic and health survey data using machine learning methods," *Diabetes and Metabolic Syndrome: Clinical Research and Reviews*, vol. 17, 2023.

[29] É. R. SANTANA, L. LOPES, D. E. MORAES, and R. Marcos, "Recognition of the Effect of Vocal Exercises by Fuzzy Triangular Naive Bayes, a Machine Learning Classifier: A Preliminary Analysis," *Journal of Voice*, 2022.

[30] Y. C. ZHANG and L. SAKHANENKO, "The naive Bayes classifier for functional data," *Stat Probab Lett*, vol. 152, pp. 137–146, 2019.